\documentclass[preprintnumbers, prd, onecolumn, floatfix,amsmath,amsmath,
preprintnumbers,showpacs, letterpaper, superscriptaddress,nofootinbib]{revtex4}
\usepackage{graphicx,epsfig}
\usepackage{amsmath}
\usepackage {amssymb}
\usepackage{subfigure}
\usepackage{bm}
\usepackage{amssymb}
\usepackage{float}
\usepackage{amsmath}
\usepackage{subfigure}
\usepackage{braket}
\usepackage[colorlinks]{hyperref}
\usepackage[usenames,dvipsnames]{color}
\hypersetup{
     breaklinks=true,
    pdfstartview={FitH},    
    colorlinks=true,       
    linkcolor=blue,          
    citecolor=red,        
    filecolor=magenta,      
    urlcolor=blue,           
    anchorcolor=green,      
    linktocpage=true
}

\def\doi{http://doi.org}

\begin{document}
\newcommand\be{\begin{equation}}
\newcommand\ee{\end{equation}}
\newcommand\bea{\begin{eqnarray}}
\newcommand\eea{\end{eqnarray}}
\newcommand\bseq{\begin{subequations}} 
\newcommand\eseq{\end{subequations}}
\newcommand\bcas{\begin{cases}}
\newcommand\ecas{\end{cases}}
\newcommand{\p}{\partial}
\newcommand{\f}{\frac}

\title{Instability and no-hair paradigm in d-dimensional charged-AdS black holes}

\author {\textbf{Afsaneh Rahmani}}\email{a$_$rahmani@sbu.ac.ir}
\affiliation{Department of Physics, Shahid Beheshti University,  Evin, Tehran, Iran}

\author {\textbf{Mohsen Khodadi}}\email{m.khodadi@ipm.ir}
\affiliation{School of Astronomy, Institute for Research in Fundamental Sciences (IPM), P.O. Box 19395-5531, Tehran, Iran}
\affiliation{Research Institute for Astronomy and Astrophysics of Maragha
(RIAAM), University of Maragheh,  P. O. Box: 55136-553, Maragha, Iran}

\author {\textbf{Matin Honardoost}}\email{m$_$honardoost@sbu.ac.ir}
\affiliation{Department of Physics, Shahid Beheshti University,  Evin, Tehran, Iran}

\author {\textbf{Hamid Reza Sepangi}}
\email{hr-sepangi@sbu.ac.ir }
\affiliation{Department of Physics, Shahid Beheshti University, Evin, Tehran, Iran}

\date{\today}

\begin{abstract}
Is it possible the no-hair paradigm is violated when a black hole undergoes
an instability? Employing massless charged scalar perturbations,
we address this question within the context of conformally invariant Einstein-Maxwell
theory for some allowed $d$-dimensional ($d=4n+4$ with conformal parameter $n=0,1,2,...$ ) topological
small AdS black-holes. We provide numerical analyses that show the non trivial
scalar hairy black hole solutions to include planar and spherical horizon topologies in higher
dimensions with an even conformal parameter $n$. It is also shown that the solutions presented here cannot be considered as scalar hairs for a
Reissner-Nordstrom background. As a result, for $d$-dimensional small AdS black holes in the presence
of a conformally invariant Maxwell source, the no-scalar hair paradigm seems to be supported.
 \end{abstract}
\maketitle

\section{Introduction}
The recent discovery of a supermassive black hole (BH) at the core of distant galaxy Messier $87^*$ by the Event Horizon Telescope (EHT) \cite{ref:eht} confirms that BHs are real astrophysical objects. LIGO \cite{ref:Ligo1, ref:Ligo2, ref:Ligo3} has already succeeded in detecting gravitational waves resulting from the merger of two BHs. These outcomes encourage one to take steps towards a more complete understanding of BH physics, not merely as  abstract objects derived from  vacuum solutions of Einstein field equations of general relativity (GR) but also as  ubiquitous objects in the universe. Interesting questions such as the faith of information trapped inside a
BH or their evaporation through various mechanisms have been raised and to some extent
answered in the past. One such interesting phenomena associated with BHs is a classical process known
as \emph{``superradiant scattering,''} through which the BH will lose energy and might eventually disappear. In other words, an incident wave can be amplified as it scatters off the BH so that the additional energy  radiated to infinity is
drawn at the expense of the BH. The event horizon,  a null one-way viscous membrane, as one of nontrivial features of a BH provides a dissipative mechanism in order to extract energy from vacuum even at the classical level.

Superradiant scattering as a radiation enhancing process is at the center of attention in various contexts. From an astrophysical point of view it is interesting since it allows
astrophysical BHs to play the role of particle detectors \cite{ref:William2017}. In recent years, interest in superradiant BHs has also grown considerably among particle physicists because it is believed that this phenomena can be used to control dark matter candidates such as axion-like particles \cite{ref:DM1, ref:DM2}. Presently, there is not a strong belief that superradiance is the wave analogue of the Penrose process for particles
\footnote{Although in both processes, we deal with energy extraction from BHs, they will occur under different conditions which indicate that the two phenomena are, in principle, separate from each other, see  \cite{ref:Pani2015} for more details.} \cite{ref:Penrose1971I, ref:Penrose1971II}
where a perturbing field with the associated charge $q$ and low-frequency $\omega$, having
possible spins $s=0,1,2$ and relevant azimuthal quantum number $m$, i.e. a scalar,
electromagnetic or gravitational wave packet, is strengthened via extracting energy
flux from a BH. Within the context of GR, this phenomenon
might happen in the small amplitude limit for a BH including the Kerr horizon with angular velocity
$\Omega_{h}$ or the Reissner-Nordstr\"{o}m (RN) horizon with  electric potential $\Phi_{h}$,
if the conditions $\omega<m\Omega_{h}$ and $\omega<q\Phi_{h}$ are met respectively \cite{ref:Pani2015}.
A very important point that should be noted is that in both types of horizons,
the relevant area/entropy does not decrease but only the rotational and Coulomb
energies are respectively extracted, see \cite{ref:Christodoulou1970, ref:Christodoulou1971, ref:Bek1973}
for more details. This means that superradiance phenomena, in agreement with our classical
intuition, will not lead to the extraction of information from a BH. Note that
in the case of formation of a mechanism that might be able to trap such superradiant modes in and around the BH, background level instability is likely to appear. Interestingly, this instability is caused by the surrounding event horizon with a perfect reflective mirror causing repeated bouncing back and forth between the mirror and horizon which could grow without bound. This growth eventually reaches
a point where the radiation pressure leads to the collapse of the mirror, sometimes known as the
``\emph{BH bomb}" \cite{ref:Zeld1971, ref:CardosoI2004, ref:Nature1972}. Concerning the Kerr BH, it
has been shown in \cite{ref:Detwe1980} that the mass of perturbing field can play the role
of a natural mirror, attracting  a considerable amount of interest on the Kerr BH instability under
a massive bosonic test field,  see \cite{ref:Kerr0, ref:Kerr1, ref:Kerr2, ref:Kerr3, ref:Kerr4, ref:Kerr5}.
However, for a RN background perturbed by a massive charged scalar field, superradiance
instability will not occur \cite{ref:Hod2013, ref:Hod2015}. In contrast, in the case of a massless
charged scalar field with small amplitude, placing a reflecting mirror  far
from the event horizon will make superradiance instability in a RN background a  possibility
\cite{ref:SHod2013, ref:Degoll2014}. Besides, given the fact that black holes are solutions
to Einstein field equations, the question arises as to what happens to such phenomena in the context
of alternative theories of gravity. Such questions have been explored and investigated in
\cite{ref:Eugeny2016, ref:Oktavio2018, ref:Nicolas2018, ref:Theodoros2018, ref:Caio2018, ref:Afsaneh2018, ref:Jia-Hui2018, ref:khodadi2020}.

Looked at from a different angle, superradiant phenomena somehow determines the
final fate of the BH involved and, as such,
can lead to the confirmation or violation of the \emph{``no-hair theorem ''} (NHT)
\cite{ref:Israel1968, ref:Carter1971, ref:Ruffini1971} which states that a perturbed BH
should eventually end in a stationary state, signified by just a small number of parameters;
mass $M$, charge $Q$ and angular momentum $J$. Historically, a series of NHTs appeared when
search was begun to look for possible interaction of BHs with any kind of matter.
An important consequence of this classical theorem is that the stable state of a BH, determined by the
above mentioned parameters, ensures that information is not lost on the initial conditions of the BH.
Despite the existence of some possibilities to violate
this theorem in models such as Einstein-Yang-Mills (EYM) \cite{ref:Volkov1989, ref:Bizon} and
Einstein-Skyrme (ES) \cite{ref:SK1, ref:SK2} with their diverse combinations with Higgs or dilaton
fields \cite{ref:DH1, ref:DH2}, it has not been conclusively proven that these models are
counter examples to NHT \cite{ref:Hertog}. It is noteworthy that the EHT data also recently opened a new window to analyze these models \cite{ref:2020}.
In \cite{ref:Stra1990}, it is shown that in the case of
non-abelian gauge fields coupled to gravity, a family of solutions with non-trivial $SU(2)$ YM
fields exist in such a way as to make  radial perturbations grow and thus becoming unstable. Concerning the Einstein-Skyrme model,
the stability condition of the Skyrme hair solution is not known at the non-linear level. In general,
it is believed that a perturbation field, such as massless/massive or a vector field etc,
once  violating the NHT,  finally results in a new stable hairy BH configuration.
As an example, the authors in \cite{ref:Dolan2015} have demonstrated that for a RN-BH, enclosed
within a cavity and perturbed by a charged, massless scalar field, it is possible to have  stable hairy solutions.
In \cite{ref:Afsaneh2018}, it has also been shown that such a possibility could exist in the framework
of alternative theories of gravity, in particular $F(R)$ gravity. However, there are some NHTs proposed
in \cite{ref:BekensteinI1972, ref:BekensteinII1972, ref:Bekenstein1995} that do not admit
such solutions. For a review involving scalar hairy solutions, see  \cite{ref:Carlos2015, ref:Carlos2020}.

A space-time well adapted to superradiance phenomena is that of anti-de Sitter (AdS) which provides
a natural, rather than an artificial reflecting boundary. Such a scenario becomes all the more interesting due
to the ``\emph{AdS/CFT correspondence}'' \cite{ref:Maldacana1998, ref:Witten1998, ref:Maldacana1999}.
The BH bomb as well as superradiant instability as two related issues, have been extensively studied within the RN-AdS and Kerr-AdS
backgrounds  \cite{ref:Cardoso2004, ref:Emanuele2009, ref:Yoshida2011, ref:Dias2014, ref:Bosch2016, ref:Oscar2017}.
It should be noted that in such a framework one is not only restricted to four-dimensional BHs. The stability of superradiant modes in higher-dimensional BHs with a RN-AdS horizon has been analytically as well as numerically studied in
\cite{ref:Oscar2012, ref:Wang2014, ref:Aliev2016, ref:Li2016, ref:Yang2017}.

In this paper we are interested in focusing on
superradiance phenomena beyond the common electrically charged-BHs which is based on the standard
Maxwell theory (SMT). Indeed, the existence of some serious shortcomings such as: vacuum polarization
in quantum electrodynamics (QED), infinite self-energy of point-like charges \cite{ref:Born1934, ref:BornII1934}
and also incompatibility with some theoretical evidence in the limit
of strong electromagnetic fields \cite{ref:Corda2018} in SMT, motivates us to embark on such a study.
Furthermore, in higher dimensions, the standard Maxwell action loses its most important property,
namely \emph{conformal invariance}\footnote{We note that in GR, due to non trivial interaction between
geometry of spacetime and matter, the presence of a matter source possessing
conformal symmetry leads to a simpler Einstein equation  \cite{ref:Mokhtar2008}.
Conformal symmetry also plays a significant role in the study of the critical phenomena in
statistical physics as well as AdS/CFT correspondence and string theory.}.
It is believed that these issues can be addressed via extending SMT to include a non-linear Maxwell modification.  One of the most popular non-linear extension of SMT is known as power Maxwell invariance
which includes a non-linear term of the form $(F_{\alpha\beta}F^{\alpha\beta})^{n+1}$ so
that if $n=\frac{d}{4}-1$ ($n=0,1,2,...$), conformal symmetry is restored \cite{ref:Mokhtar2008, ref:Mokhtar2007, ref:Hendi2009, ref:Rasegar2009, ref:Ahmad2012}. Note that the underlying $d$-dimensional framework has two clear differences
compared to the standard higher-dimensional counterpart. First, unlike the standard case, gravity is
coupled to a non-linear Maxwell field. Second, the relevant electrodynamic source
is a \emph{``conformally invariant Maxwell source''} (CIMS).

Overall, our main purpose in this work is to evaluate the validity of NH paradigm in higher
dimensions by investigating instability of a non-standard
$d$-dimensional small RN-AdS BH, that is $r_{h}\ll L$ where $r_{h}$ and $L$
are the event horizon and AdS radius, perturbed by a charged, massless scalar field. More precisely, we deal with a new class of $d$-dimensional RN-AdS BH within the framework of a conformally
invariant Einstein-Maxwell gravity. We use numerical solutions to argue that there is the possibility
of supporting the NH conjecture by CIMS in higher dimensions allowed by the theory.

\section{Scalar hairy black holes in $d$-dimensions with a CIMS source \label{BH}}

\subsection{ The setup}

We consider Einstein gravity in a $d$-dimensional AdS spacetime coupled to a CIMS source as well
as a charged, massless scalar field, described by the action
\begin{eqnarray}\label{action}
S[g_{\alpha\beta},A_{\alpha},\Psi,\Lambda]=\int d^{d}x\sqrt{-g}\bigg( R
-2\Lambda -\frac{1}{4} (F^{\alpha\beta }F_{\alpha\beta })^{d/4}-\frac{1}{2}g^{\alpha\beta}(D_{(\alpha}^* \Psi^* D_{\beta)} \Psi)\bigg)~,
\end{eqnarray}
with $F_{\alpha\beta}=\nabla_\alpha A_\beta-\nabla_\beta A_\alpha$ and $D_\alpha=\nabla_\alpha-iqA_{\alpha}$ where for $d=4$, the familiar Einstein-Maxwell action
coupled to a scalar field $\Psi$ is recovered. In the above action, $\Lambda $ represents an AdS spacetime with a negative value given by
$-(d-1)(d-2)/2L^{2}$ for asymptotic solutions with $R$ and $L$ respectively denoting a scalar curvature and length scale. It is well known that the electro-magnetic part of the above action  $S[A_{\alpha}]$ is invariant under conformal transformation $g_{\alpha\beta}\rightarrow\Omega^2 g_{\alpha\beta}$ and $A_\alpha\rightarrow
A_\alpha$.  Also the conformal invariance of $S[A_{\alpha}]$ is guaranteed via the traceless term $T_{\,\,\,\alpha}^{\alpha}=0$ for energy-momentum
tensor which will only occur for $d=4n+4$ with the non-linear parameter $n=0,1,2,...$ \cite{ref:Mokhtar2008, ref:Mokhtar2007, ref:Hendi2009, ref:Rasegar2009, ref:Ahmad2012}. By varying the above action with respect to $g_{\mu
\nu }$, the gauge field $A_{\mu }$ and scalar field $\Psi$, the following field equations are derived
\begin{eqnarray}
&&G_{\alpha\beta }+\Lambda g_{\alpha\beta }=T_{\alpha\beta}=(T_{\alpha\beta}^{EM}+T_{\alpha\beta}^{\Psi})~,\label{EoM1}\\
&&\frac{1}{\sqrt{-g}}\partial _{\alpha}\bigg( \sqrt{-g}F^{\alpha\beta }(F^{\alpha\beta }F_{\alpha\beta })^{n}\bigg) =J^{\beta}~,\label{EoM2}\\
&&D_{\alpha}D^{\alpha}\Psi=0~, \label{EoM3}
\end{eqnarray}
with
\begin{eqnarray}
&&T_{\alpha\beta}^{EM}=(n+1)F_{\alpha \sigma }F_{\beta
}^{~\sigma }(F^{\alpha\beta }F_{\alpha\beta })^{n}-\frac{1}{4}g_{\alpha\beta }(F^{\alpha\beta }F_{\alpha\beta })^{n+1}~,\label{TF}\\
&&T_{\alpha\beta}^{\Psi}=D^{*}_{(\alpha}\Psi^{*}D_{\beta)}\Psi-\frac{1}{2}g_{\alpha\beta}\big(g^{\rho \sigma}D^{*}_{(\rho}\Psi^{*} D_{\sigma)}\Psi\big)~,
\label{TS}\\
&&J^{\beta}=\frac{iq}{2}\bigg(\Psi^*D^{\beta}\Psi-\Psi(D^{\beta}\Psi)^*\bigg),\label{J}
\end{eqnarray}
where $\nabla_\beta J^\beta=0=\nabla_{\alpha}T^{\alpha\beta}$. Note that due to $U(1)$ symmetry in the underlying theory, the scalar field $\Psi$ as well as vector potential $A_\alpha$ enjoy \emph{``gauge freedom''} which keeps $F_{\alpha\beta}$
and $D_\alpha{\Psi}$ invariant under the following gauge transformations
\begin{eqnarray}\label{gauge}
\Psi \rightarrow \exp({i \chi})\Psi,~~~~ A_\alpha \rightarrow A_\alpha+q^{-1} \chi_{,\alpha}~.
\end{eqnarray}
Here, $\chi$ can be any real scalar field. Gauge freedom, in essence, allows us to work in any cost-effective gauge.
In subsection C, one will see that the existence of this property which would enable us to extract static solutions in the non-linear regime, is very essential \cite{ref:Dolan2015}.

Neglecting the effects of scalar field in test approximation, the underlying theory described by (\ref{action}) admits a spherically symmetric BH solution as follows
\begin{equation}\label{met}
ds^{2}=-\eta(r)dt^{2}+\frac{dr^{2}}{\eta(r)}+r^{2}d\Omega _{4n+2}^{2},
\end{equation}%
with the laps function $\eta(r)$, the line element $d\Omega _{4n+2}^{2}$
of a $(4n+2)$-dimensional hypersurface and curvature constant $k$ defined as
\[
d\Omega _{4n+2}^{2}=\left\{
\begin{array}{cc}\label{line}
d\theta _{1}^{2}+\sum\limits_{i=2}^{4n+2}\prod\limits_{j=1}^{i-1}\sin
^{2}\theta _{j}d\theta _{i}^{2} & k=1 \\
\sum\limits_{i=1}^{4n+2}dx _{i}^{2} & k=0\\
d\theta _{1}^{2}+\sinh ^{2}\theta _{1}d\theta _{2}^{2}+\sinh ^{2}\theta
_{1}\sum\limits_{i=3}^{4n+2}\prod\limits_{j=2}^{i-1}\sin ^{2}\theta
_{j}d\theta _{i}^{2} & k=-1%
\end{array}%
\right.
\]%
Interestingly, the CIMS equation (\ref{EoM2}) in the absence of the scalar field
implies that the electric field in $(4n+4)$-dimensions  still obeys the inverse-square law
i.e. $F_{tr}=\frac{Q}{r^{2}}$. In other words, in the presence of a conformally invariant
source the relevant expression of the electric field is the same as that for
the RN solution in $4d$, independent of the dimension. It is worth noticing that the root
of this feature comes from the CIMS so that in the framework of a SMT based higher-dimensional
theory, the electric field is dependent on the dimension. As a result, within the extended CIM
setup and based on vanishing electromagnetic potential at the event horizon $r_{h}$,
we may take the following anzats for the gauge potential
\begin{equation}\label{guage}
A_\alpha dx^\alpha=\big(\phi(r)dt,0,0,0\big)\Rightarrow A_{\alpha}=\left(\frac{-Q}{r}+C, 0, 0,0 \right),
\end{equation}
similar to its 4$d$ counterpart. It is worth mentioning that the constant of integral $C$ comes from AdS/CFT correspondence which would allow us to make the vector potential vanish on the event horizon, i.e. $A_{\alpha}(r_{h})=0$ \cite{ref:Horo2011,
ref:Herzog2009}. As is shown in detail in \cite{ref:Yoshida2011}, the constant $C$ only shifts the real part of the frequency, without any effect on the imaginary part, which is an essential characteristic in studying superradiant instability. In other words, $U(1)$-gauge transformation (\ref{gauge}) guarantees that presence of $C$ just makes the scalar field $\Psi$ to undergo a frequency shift as $\mbox{exp}(iqCt)\Psi$ \cite{ref:bosch2019}.
Therefore, to study superradiance phenomena, one takes $C=0$ in the linear approximation, just as was done in \cite{ref:Yoshida2011, ref:Wang2014, ref:bosch2019}. However, there is a point  worth paying attention to; the solution for the scalar field in the background for which $A_\alpha(r_h) = 0$ does not represent the same state as the one in the background that satisfies $A_\alpha(r_h) \neq 0$ since they are related through a $U(1)$ transformation which does not vanish at infinity. So these two backgrounds actually address states that obey different boundary conditions.
This will become clear in the next subsection.

Now, taking the $(rr)$ component in Eq. (\ref{EoM1}), the lapse function
$\eta(r)$ reads
\begin{equation}\label{f}
\eta(r)=k-\frac{m}{r^{4n+1}}+\frac{2^{n}Q^{2n+2}}{r^{4n+2}}+\frac{r^{2}}{L^{2}}~,
\end{equation}
which can be written in a more useful form
\begin{equation}\label{fr}
\eta(r)=k-\frac{2m(r)}{r^{4n+1}}+\frac{r^{2}}{L^{2}}~,~~~
\mbox{with}~~~ m(r)=m+\frac{2^{n-1}Q^{2n+2}}{r}~,
\end{equation}
were, $m$ is an integration constant representing
the mass of BH. Note that the above laps function $\eta(r)$ is quite
different from that obtained in a higher-dimensional SMT-based space since,
as is clear, the electric charge term goes as $ r^{-(4n+2)}$ while
the standard Maxwell source gives $r^{-(8n+2)}$. This means that the field arising
from a CIMS falls off  slower than that given by a
standard Maxwell source. In the lapse function (\ref{f}), $k=1, 0, -1$ represent solutions with spherical, planar and
hyperbolic geometries for the horizon, respectively. In what
follows, we will see the important role played by  the topology of the horizon
in describing the superraidiance phenomena.

\subsection{Scalar wave scattering: linear regime \label{BHs}}
Let us now focus on the scattering of a scalar test field
in a $d$-dimensional BH background with a CIMS. This will be done
at the linear level which means that having a small amplitude for scalar field,
we can ignore the changes induced by its fluctuations so that
the spacetime geometry can still be considered as fixed.
The spherical symmetry characteristic of the test
scalar field $\Psi$, suggest the following perturbation anzats
\begin{equation}\label{psi}
\Psi= \frac{\psi(r)}{r^{2n+1}}\exp(-i\omega t),
\end{equation}
where $\psi(r)$ is the radial part of the wave function and $\omega$ is the frequency. Inserting the above expression in Klein-Gordon
equation (\ref{EoM3}), in the presence of gauge potential (\ref{guage}),
we arrive at the following equation
\begin{equation}\label{Sch0}
\eta^{2} \psi^{''}+\eta \eta^{'}\psi^{'}+U_{eff}\psi=0,
\end{equation}
with
\begin{equation}\label{Sch1}
U_{eff}=\big(\omega+qA_{t}\big)^2-\bigg(\frac{n(4n+2)}{r^2}\eta^2+\frac{2n+1}{r}\eta \eta^{'}\bigg),
\end{equation}
where a prime denotes  derivative with respect to $r$. Using tortoise coordinate $r_{*}$,  $\eta^{-1}=\frac{dr_*}{dr}$,
equation (\ref{Sch0}) is re-expressed as a Schrodinger-like equation
\begin{equation}\label{Sch2}
\frac{d^2\psi}{dr_{*}^2}+U_{eff}\psi=0~,
\end{equation}
the potential $U_{eff}$, depending on the model at hand,
describes the test field as well as the curvature of the background.
For $r$ going to spatial infinity or the event horizon, the behavior of the radial wave function in these limits are
\begin{eqnarray}\label{limit}
 && \psi(r)\sim \mbox{exp}\big(-i(\tilde{\omega}-\frac{qQ}{r_h}) r_{*}\big), \qquad  \tilde{\omega}\equiv\omega+qC\qquad r \rightarrow r_{h} ,\nonumber\\
 && \psi(r)\sim r^{-(2n+2)}  \qquad  \qquad \qquad \qquad  \qquad \qquad \qquad ~~~ r \rightarrow +\infty,
\end{eqnarray}
where the condition of no outgoing wave from the event horizon has been imposed.
Here, it is interesting to note that
invariance of equation (\ref{Sch0}) under
transformations $q\rightarrow-q$ and  $Q\rightarrow-Q$ means that, without loss of generality,
we may fix the positive sign for $qQ$.

 There is a subtlety regarding the above argument which merits a more detailed discussion at this point. One  way to realize superradaince phenomena is by studying the energies as well as propagation of the scalar wave function.
To have superradiance, it is necessary that signs of the group and phase velocities of the scalar field wave function become negative and positive respectively, that is $v_g<0$ and $v_p>0$. Also, the $(tr)$ component of the energy-momentum tensor should be nonzero.
In what follows, by taking the gauge $C=0$, we investigate the above statements to see if they are satisfied. Regarding the radial solution (\ref{limit}), one finds that the behavior of the perturbed scalar field near the horizon is as follows
\begin{equation}\label{limit1}
\Psi= \frac{\exp^{-i\omega t}\exp^{ikr_*}}{r^{2n+1}}\psi(r),
\end{equation}
where $k=-(\omega+q A_t(r_h))$ and $A_t(r_h)=-\frac{Q}{r_h}$.
This means that the group velocity is $v_g=\frac{d\omega}{dk}=-1$,
implying that information is passing into the BH which is in agreement with the boundary condition demanded at the horizon. As for the phase velocity, we also have
\begin{equation}\label{velocity}
v_p=\frac{\omega}{k}=\frac{\omega}{-\big(\omega+q A_t(r_h)\big)}>0.
\end{equation}
Here, the positivity of $v_p$ is guaranteed since in a superradiant regime, $\omega<-q A_t(r_h)$.

Positive phase velocity means that energy is extracted off the BH
from the point of view of a horizon observer. From the near-horizon behavior of the scalar field,
$\Psi(r)\propto \exp^{-i \omega t}\exp^{-i(\omega+q A_t(r_h))r_{*}} $, a distant observer
also recognizes  that the wave is coming out of the BH provided that the superradiant condition,
$\omega<-q A_t(r_h)$, is satisfied.
In order to investigate the third quantity mentioned above, we calculate the $(tr)$ component of the
energy-momentum tensor and find $$T_{tr}\propto -\frac{(\omega+q A_{t}(r_h)) k}{\eta}=\frac{k^2}{\eta}>0,$$
meaning that the energy flux at the horizon is non-zero.

As an alternative approach to the above arguments, one may study the effects of the scalar perturbation on energy extraction from the BH.
To do so, we consider the conserved energy flux for the scalar field, $J_{\mu} = - T_{\mu \nu} \xi^{\nu} $, using the energy-momentum tensor $T_{\mu \nu}$ and Killing vector $\xi^{\nu}$,
which is timelike at infinity \cite{ref:wald}.
By projecting the energy flux vector along the null horizon to generate the Killing vector
of the BH background $\chi^\mu$, we obtain the following rate of energy extraction from the BH
(further details can be found in \cite{ref:fermion})
\begin{eqnarray}\label{1}
-\langle J_\mu \chi^\mu\rangle=\langle J_\mu n^\mu\rangle=\int dA n^\mu J_\mu\,,
\end{eqnarray} where $n^\mu=-\chi^\mu$ is the appropriately directed normal to the horizon.
Therefore, due to continuity and stationarity of the scalar field, the energy flux transferred
to the horizon should be carried away by a conserved radial current $J_r=i(\Psi^{*}\Psi_{;r}-\Psi\Psi_{;r}^{*})$
so that the integral of $\mathcal{F} \equiv \int dA n_\mu J^\mu$ over the horizon represents the net radial energy flux extracted from the BH. Here, $n_\mu\sim \nabla_r r$ represents a radial normal vector.
It is now clear that for superradiance to occur, i.e. having outward energy flow,
it is necessary that $n_\mu J^\mu>0$. Now for the complex scalar field perturbation near the
horizon, given by (\ref{limit1}), we have
\begin{equation}\label{current2}
n_{r}J^{r}=\frac{2k}{r^{4n+2}}|\psi(r)|^2 .
\end{equation}
It is clear that $n_r J^r>0$ provided that $k>0$.
This is not possible unless the superradiant condition ($\omega<\frac{q Q}{r_h}$) is met.
Given the above arguments, one interesting point to note is that by transforming the scalar solution obtained in the gauge $A_t(r_h)\neq0$ ($C=0$) to gauge $A_t(r_h)=0$ ($C\neq0$), the scalar field acquires a frequency shift that results in phase and group velocities with opposite signs on the horizon\footnote{When one imposes that the vector potential to vanish at the event horizon i.e. $A_{t}(r_{h})=0$, it means that the scalar filed will undergo a frequency shift according to $e^{i\frac{qQ}{r_{h}}t}$. The behavior of the scalar field perturbation near the event horizon is thus given by
		\begin{equation}\label{psi2}
		\Psi= \frac{e^{ikr_*}}{r^{2n+1}}e^{-i(\omega-\frac{qQ}{r_{h}}) t} \qquad   r \rightarrow r_{h},
		\end{equation}
		where $k=-\omega$. So the group velocity is $v_g=\frac{d\omega}{dk}=-1$ while the phase velocity is positive
		\begin{equation}\label{velocity}
		v_p=\frac{\omega-\frac{qQ}{r_{h}}}{k}=\frac{\omega-\frac{qQ}{r_{h}}}{-\omega}>0,
		\end{equation}
		when the superradiance condition, $\omega<\frac{qQ}{r_{h}}$, is satisfied. This clearly shows that the sign of the group and phase velocities can be different even when one considers $A_{\alpha}(r_{h})=0$, so that the energy will be extracted off the black hole for this choice of the gauge.}. However, without resorting to a $U(1)$ transformation, {\it i.e.} if from the beginning one  works in the gauge $A_{\alpha}(r_{h})=0$, a frequency shift will no longer appear and as a result, the sign of phase and group velocities on the horizon will be the same. As a consequence, the background with $A_{\alpha}(r_{h})=0$ actually is free of superradiant instability. However, through numerical analysis, it will be shown below that the underlying background still undergoes an instability.

\begin{figure}[!ht]
\includegraphics[width=8cm,height=5.5cm]{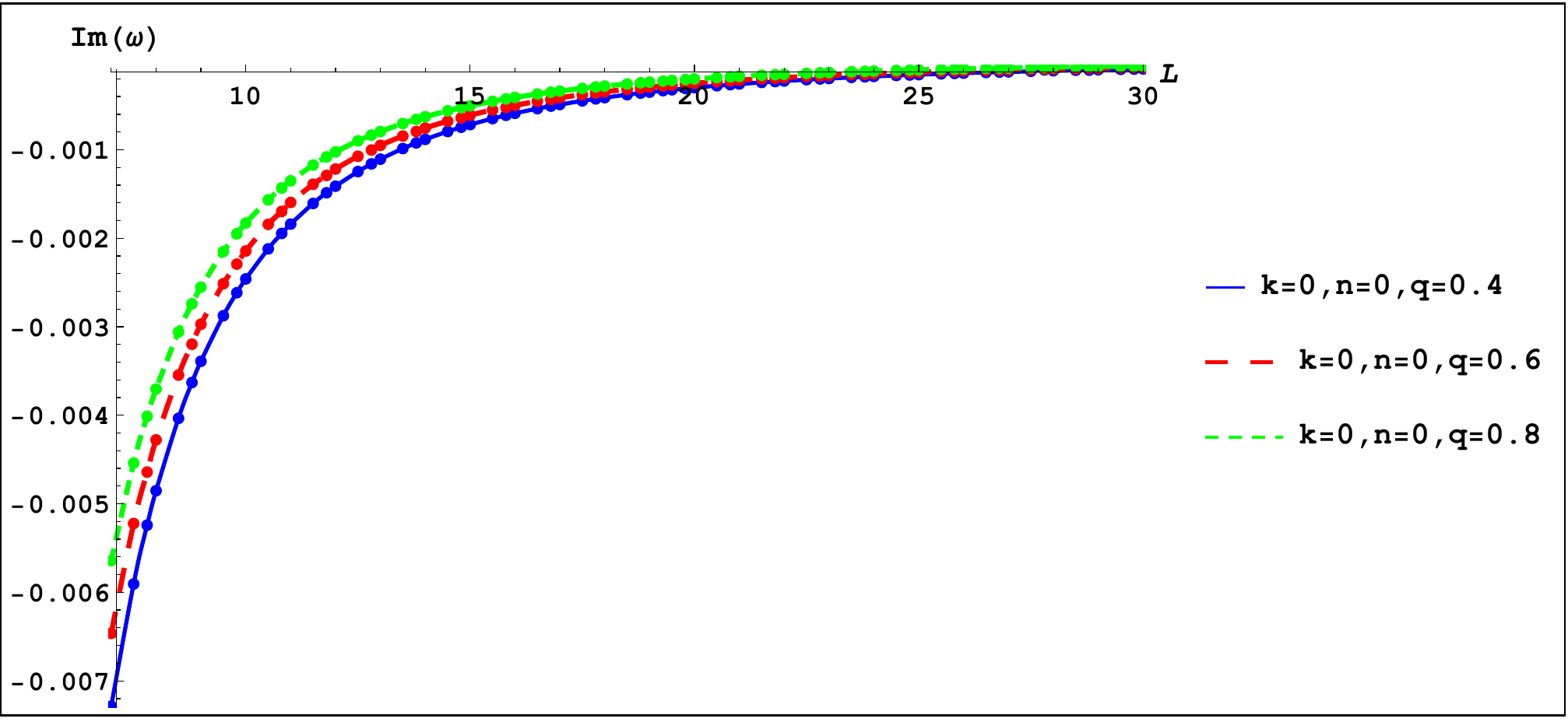}
\includegraphics[width=8cm,height=5.5cm]{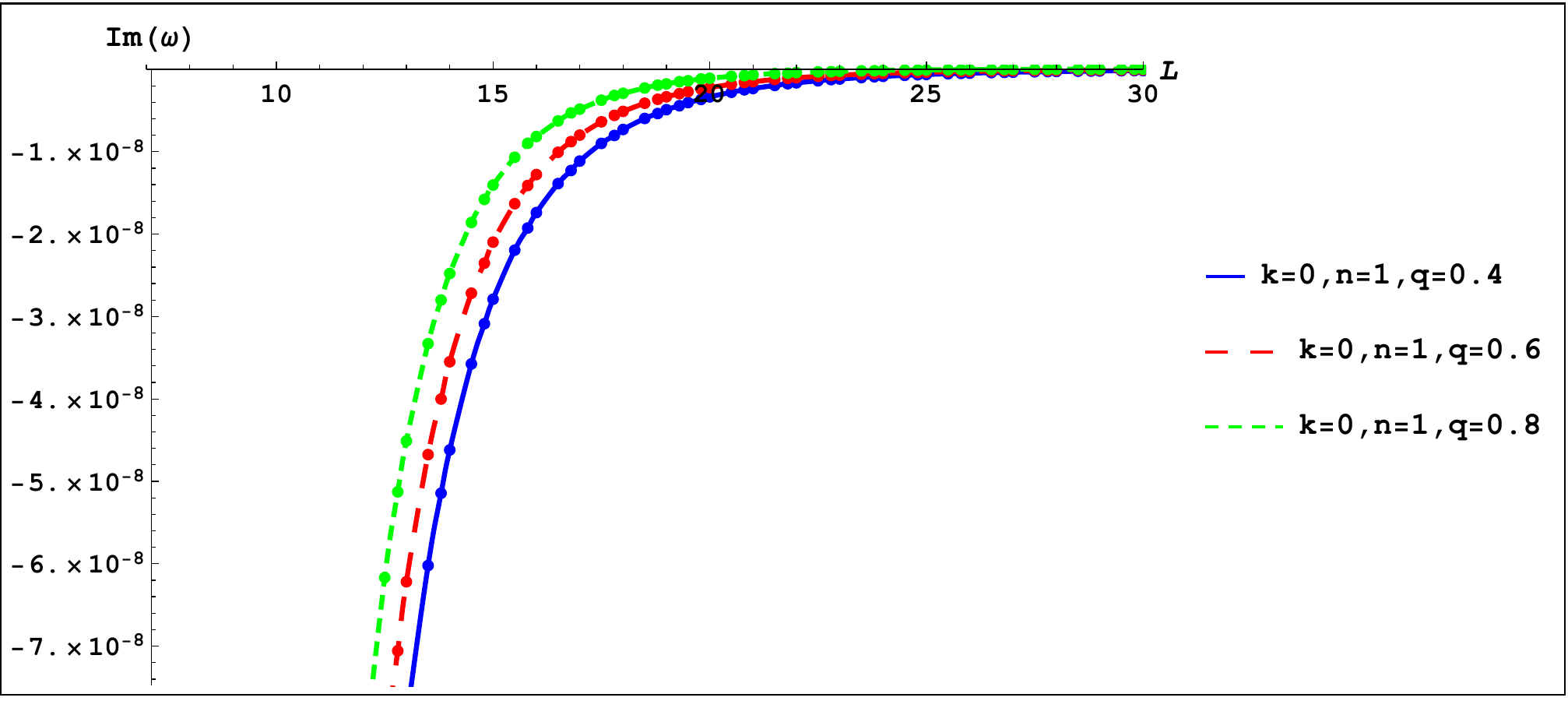}\\
\includegraphics[width=8cm,height=5.5cm]{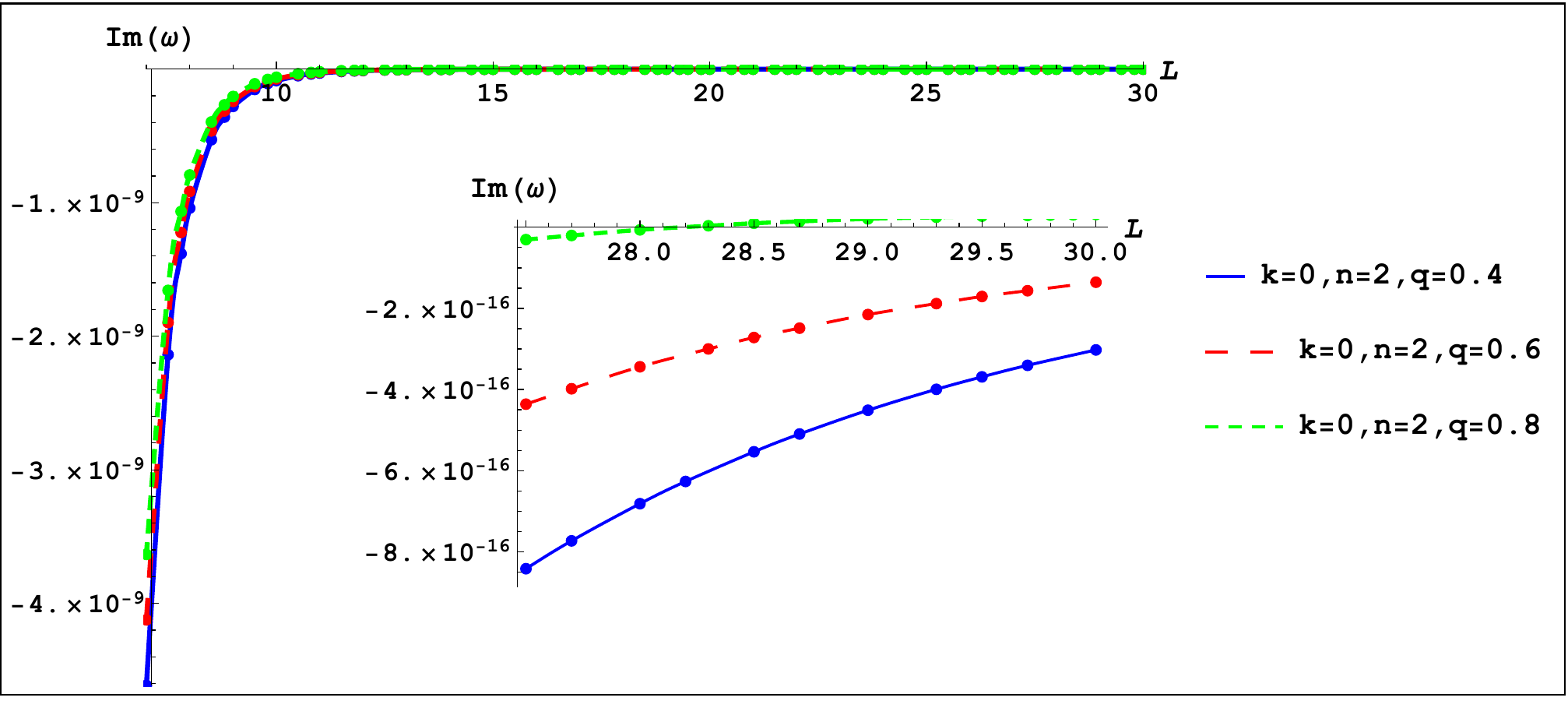}
\includegraphics[width=8cm,height=5.5cm]{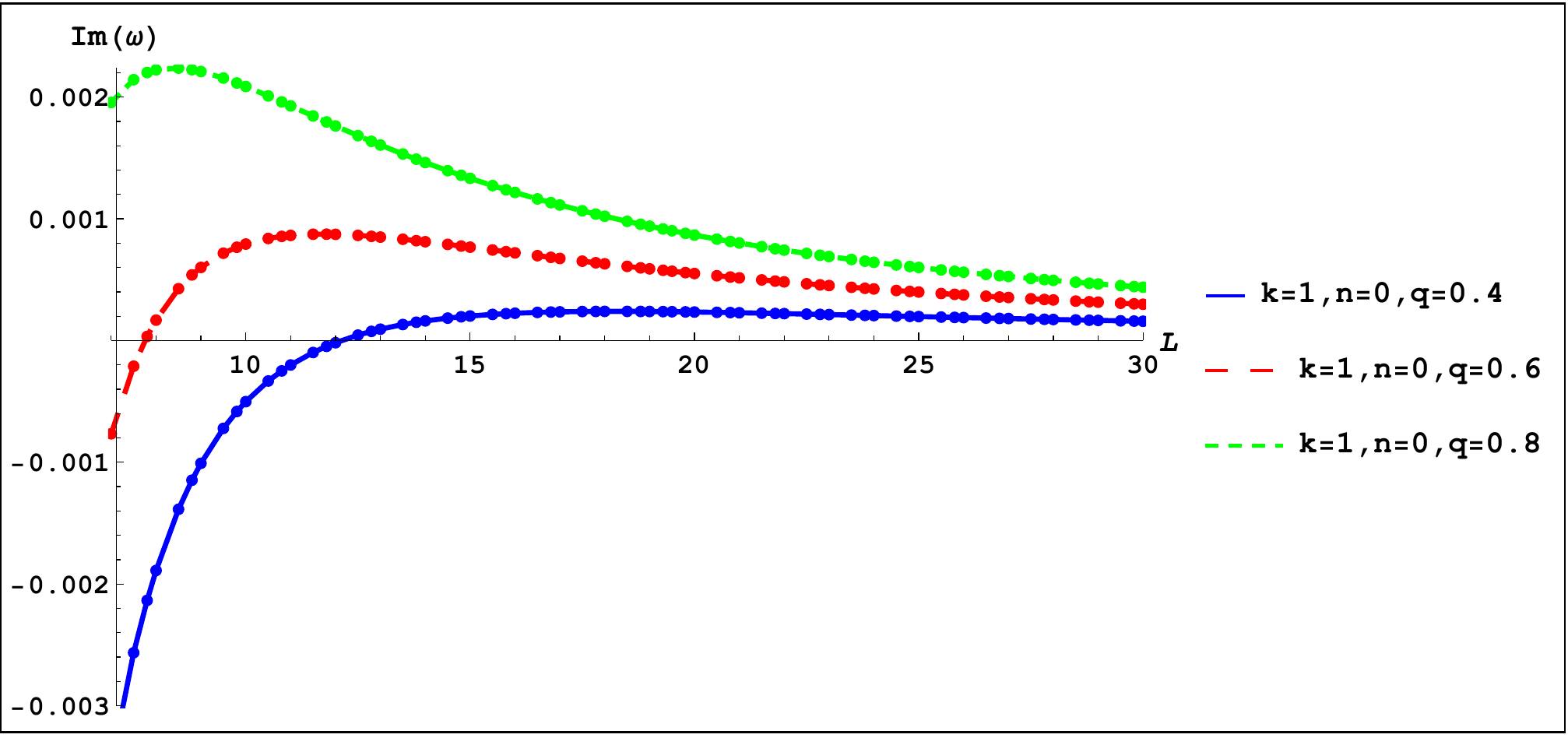}\\
\includegraphics[width=8cm,height=5.5cm]{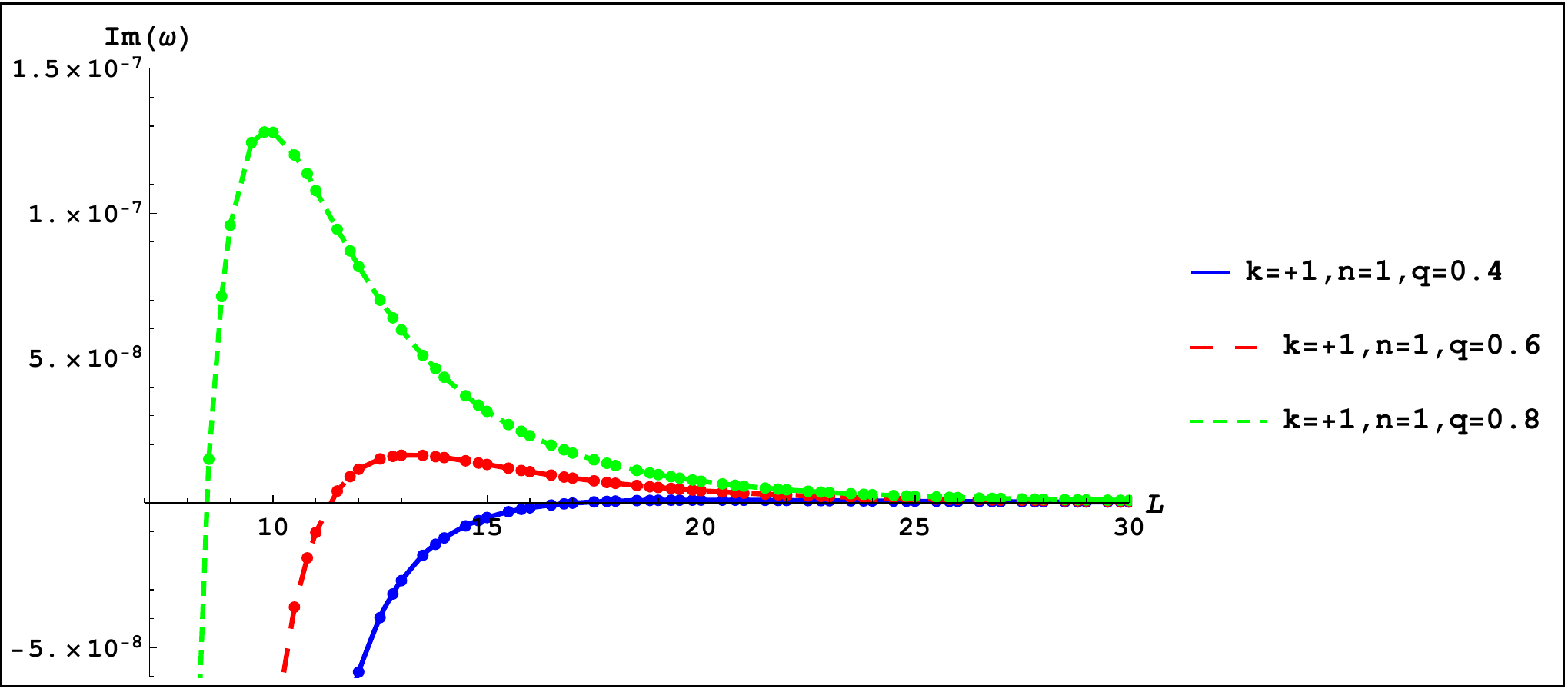}
\includegraphics[width=8cm,height=5.5cm]{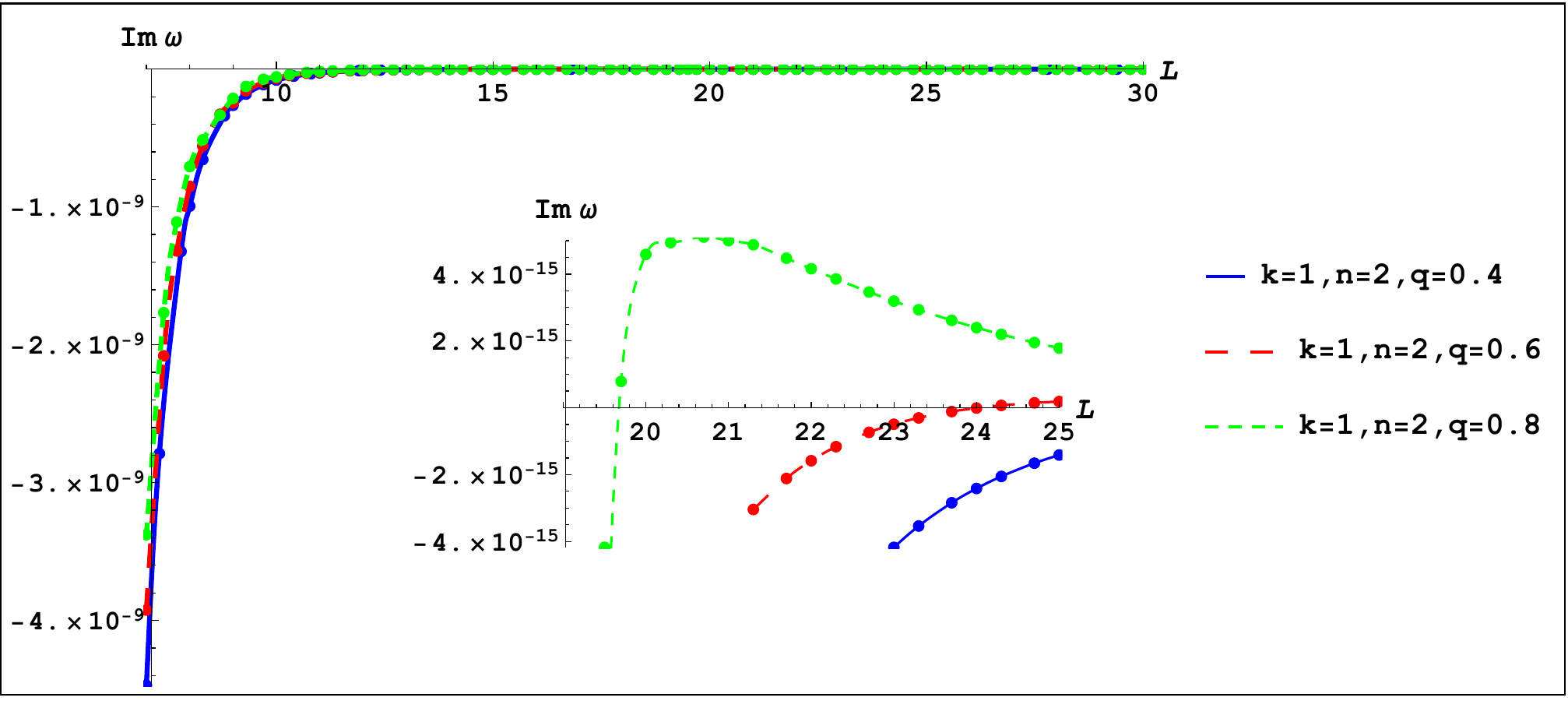}
\caption{The imaginary part of the frequency plotted in terms of the length of AdS for different values
of conformal parameter $n=0,1,2$ (d=4, 8, 12) as well as different curvatures of the event horizon ($k=0,k=+1$).
We set the values
$Q=0.9$ and $m=1$ for charge and mass of BH, respectively.}
\label{Super}
\end{figure}

By imposing a natural reflecting
AdS boundary that guaranties the vanishing of the scalar test field,
$\psi(r=L)=0$, we will deal with a two-point boundary value problem
that results in a discrete spectrum for $\psi(r)$ in (\ref{Sch0}) with
frequency modes $\omega_m$ having real and imaginary parts, $\omega_{m}=\omega_{R}+i\omega_{I}$.
Here, the label $m$ denotes the number of nodes related to scalar field modes trapped in
the region $r_h<r<L$. Given the fact that the relevant anzats for the scalar field (\ref{psi}) is time
dependent, it is clear that if $\omega_I > 0$ the amplitude grows exponentially
and then in the background instability occurs. However
if $\omega_I <0$, then due to exponential decay of the amplitude of the scalar
field modes, instability will be absent. Note that the validity of these
statements is independent of the sign of $\omega_R$ as well as the dimension of the space-time
so that without loss of generality it can be applicable to higher-dimensional BHs.
As a result, $\omega_I$ is the parameter which determines the realization
of instability. Therefore, focusing on specifying the sign of $\omega_I$,
in the rest of this subsection we will investigate the possibility of the realization of instability
for RN-AdS BHs in the framework of a conformally invariant Einstein-Maxwell  theory.
Following the method adopted in \cite{ref:Dolan2015}, via integrating the radial
perturbation equation (\ref{Sch2}) from $r=r_h(1+\epsilon)$ to $r=L$ and fixing an
initial value for $\omega_m$\footnote{{Using a matching method for near and far region solutions of the radial equation (\ref{Sch2}) for small BHs, the authors in \cite{ref:Yoshida2011,ref:Wang2014} analytically found the frequency  $\omega_{m}$. Changing the gauge potential in \cite{ref:Wang2014} to (\ref{guage}), we used the results of \cite{ref:Wang2014} to fixe the initial value for $\omega_{m}$. }} we numerically compute the spectrum of modes
for the underlying BHs with different horizon topologies.
Fig. \ref{Super} is a summary of the results obtained from the numerical
solution. First, the numerical solution for the event horizon with
open geometry ($k=-1$) is not favored since for this topology the small BH approximation
($r_{h}\ll L$) is not met.
For  planar topology  ($k=0$), the numerical
solution shows that by increasing the value of the conformal parameter $n$ as well as scalar charge $q$, the probability of
occurrence of instability increases. However, as is clear from Fig. \ref{Super}, for $n=0,1$
($d=4,8$), increasing the scalar charge $q$, we have $\omega_I\leq0$ but for $n=2$, the sign of $\omega_I$
at $L=28$ with $q=0.8$ crosses over to positive values\footnote{Concerning the planar AdS horizon in four dimensions ($n=0$), superradiant instability is absent and instead we deal with another type of instability related to an AdS background, known as ``near-horizon scalar condensation instability.'' Unlike superradiance instability, this type of instability is suppressed in small BH approximation. A four dimensional AdS background with planar horizon does not satisfy small horizon
approximation $r_h \ll L $ and therefore superradiant instability is absent while the near-horizon scalar condensation instability is present, see \cite{ ref:Oscar2017,ref:SS2008} for more details. Our numerical analysis however shows that this cannot be extended to $n>0$ since for
$n=1,2$ small BH approximation is restored, so that the near-horizon scalar condensation instability is suppressed and we only deal with superradiant instability.}. This may be interpreted as having an enhanced chance of occurrence of
superradiance for large values of the scalar charge and AdS radius in $d$-dimensional AdS small BHs with flat horizon sourced by a CIM.
Concerning the BHs with spherical geometry for the horizon ($k=+1$), Fig. \ref{Super} tells us that
despite the fact that by increasing the scalar charge $q$, the instability becomes stronger,
going to allowed higher dimensions, its probability of occurrence may become weaker.
In other words, according to $\tau = \frac{1}{\omega_{I}} $ one can see by decreasing the value of $\omega_{I}$, a longer time is required to achieve instability. Also we see that in allowed higher
dimensions the starting point of instability ($\omega_I\sim0$ corresponding to very long-lived
quasi-normal modes \cite{ref:Deg,ref:Rich}) is located at distances away from the
event horizon ($r_h$). So, like in the previous case, here again increasing the scalar charge
in higher dimensions allowed by CIMS, the instability is realizable.
Generally,
by employing a numerical method and a linear approximation, we have so far shown that scattering of a scalar test field off a $d$-dimensional small BH (with topological indexes $k=0,+1$ as well as a CIMS), will lead to instability. As a supplementary point, one may regard the instability discussed above as a superconducting-like instability since it is shown that the gravitational dual of a holographic superconducting  system
is acquired via the coupling of AdS-gravity to a system consisting of Maxwell and charged scalar fields \cite{ref:Hartnoll2008}.
In other words, the superconducting phase transition is actually equivalent to a classical instability of an AdS-BH perturbed by a charged scalar field \cite{ref:Cardoso2014}.

\subsection{Static hairy solutions: non-linear regime}
So far, using a standard spherically symmetric metric (\ref{met}), we have
studied the realization of instability at leading order, i.e. in a fixed background.
However, in what follows, by going beyond linear approximation, we may extract a family of
static BH solutions including charged scalar-field hair which is surrounded by the event horizon and
a natural AdS boundary. To do so, we generalize metric (\ref{met}) to
\begin{equation}\label{metI}
ds^{2}=-\eta(r)\xi(r)dt^{2}+\frac{dr^{2}}{\eta(r)}+r^{2}d\Omega _{4n+2}^{2},
\end{equation}
which includes an additional lapse function $\xi(r)$ carrying the
matter field backreaction effects on the spacetime geometry \cite{ref:Dolan2015}.
 Before going any further, it should be noted that ansats (\ref{psi}) represents
the fact that the scalar perturbations near critical frequency are not static
but time dependent. By appealing to $U(1)$ symmetry one is able to extract static scalar hair solutions from time-dependent
ansatz (\ref{psi}) at the non-linear level. Specifically speaking, the static scalar hair solutions are commonly obtain by using  $\chi=\omega t$ in gauge transformation (\ref{gauge}) \cite{ref:Dolan2015}. This eliminates the time dependency of the scalar field as  $\Psi\equiv\psi(r)$ and the time component of vector partial $A_t$ becomes
\begin{eqnarray}\label{gauge2}
A_t = -\frac{Q}{r}+\frac{\omega}{q}~.
\end{eqnarray} As a result, at the critical frequency $\omega_c=\frac{qQ}{r_h}$,  we automatically have $A(r_h)=0$.  Therefore, the gauge transformation (\ref{gauge}) will render $A_t$ static, making the search for hairy static solutions near critical frequency in which the scalar field and BH are in equilibrium, a possibility. For $\omega<\omega_c$, the system gets out of equilibrium and makes the occurrence of instability possible \cite{ref:Dolan2015}. One of the reasons for demanding $A(r_h)=0$ in studying instability and extracting scalar hairy BH solutions, is the requirement of having a regular event horizon \cite{ref:Dolan2015}.

Now by introducing the gauge potential $A_t \equiv A_0(r)$ due to gauge transformation (\ref{gauge2}) and fixing the generalized metric (\ref{metI}),
the equations of motion (\ref{EoM1})-(\ref{EoM3}) result in the following four equations
\begin{eqnarray}
&&\eta(r)^2(1+2n)\xi'(r)-q^2 r A_{0}(r)^2 \psi(r)^2-\eta(r)^2 r\xi(r) \psi'(r)^2=0~, \label{static1}\\
&&\eta^{^{\prime }}(r)+\frac{4n+1}{r}\big( \eta(r)-k\big) +\dfrac{2^{n-1}}{\xi(r)^{n+1}}rA_{0}^{'}(r)^{2n+2}-
\frac{(4n+3)}{L^2}r+\frac{\xi'(r) \eta(r)}{2 \xi(r)}=0~,  \label{sratic2}\\
&&\big(2n+1\big)\eta(r)A_{0}^{'}(r)^{2n}A_{0}^{''}(r)+\left(\frac{(4n+2)\eta(r)}{r}-
\dfrac{\eta(r)\xi'(r)}{2\xi(r)}\right)A_{0}^{'}(r)^{2n+1}-(-2)^{-n}q^2\xi(r)^{n}\psi(r)^2A_{0}(r)=0~,\label{static3}\\
&&\eta(r)\psi^{''}(r)+\bigg(\eta^{'}(r)+\frac{(4n+2)\eta(r)}{r}+\dfrac{\eta(r)\xi'(r)}{2\xi(r)}\bigg)\psi^{'}(r)+
\frac{q^2A_{0}(r)^2 \psi(r)}{\eta(r)\xi(r)}=0~, \label{static4}
\end{eqnarray}
where $A_{0}^{'}(r)=E$ with $E$ being the electric field. It seems that this set of equations cannot be solved analytically so
in what follows, we numerically provide a possible family of solutions for the above coupled equations.

Let us first discuss the boundary conditions on the BH horizon, $r = r_h$ and the AdS boundary,
$r=L$. Assuming regular horizon, that is, $\eta(r_h)=0$ and $\eta^{'}(r_h)>0$
we have $m(r_{h})\equiv m_{h}=\frac{r_{h}^{4n+1}}{2}(k+\frac{r_{h}^2}{L^2})$. For the value of the additional laps function $\xi (r)$ at the event horizon we set $\xi(r_{h})=1$ while the requirement to have finite physical quantities on the horizon
imposes the conditions $A(r_{h})\equiv A_{h}=0= \psi^{'}_h\equiv\psi^{'}(r_h)$  for the field variables. As a result, employing a Taylor series expansion
\begin{eqnarray}
&&m=m_{h}+m'_{h}(r-r_{h})+...,  \label{Taylor1} \\
&&A_{0}=A_{h}+A'_{h}(r-r_{h})+\frac{A''_{h}}{2}(r-r_{h})^2+...,  \label{Taylor2} \\
&&\psi=\psi_{h}+\psi'_{h}(r-r_{h})+\frac{\psi''_{h}}{2}(r-r_{h})^2+..., \label{Taylor3}
\end{eqnarray}
for the field variables in the near event horizon and substituting them
into field equations (\ref{static1})-(\ref{static4}), we arrive at
\begin{eqnarray}
&&m'_{h}=2^{n-2}r_{h}^{4n+2}E_h^{2n+2}~, \label{Taylor4} \\
&&\xi'_{h}=\frac{r_{h}^{3} q^{2} \psi_{h}^{2} E_{h}^{2}}{(1+2n)\left(-2^{n-1}r_{h}^{2}E_{h}^{2n+2} +(4n+1)k+\frac{(4n+3)r_{h}^{2}}{L^2}\right)^2}~, \label{Taylor4} \\
&&A^{''}_h=\frac{(-2)^{-n} q^{2}\psi_{h}^{2}{E_{h}}^{1-2n}r_{h}}{(2n+1)\left(-2^{n-1}r_{h}^{2}E_{h}^{2n+2} +(4n+1)k+\frac{(4n+3)r_{h}^{2}}{L^2}\right)}+\left(\frac{\xi'_{h}}{2}-\frac{4n+2}{r_{h}}\right)\frac{E_{h}}{2n+1}~,  \label{Taylor5} \\
&&\psi^{''}_h=-\frac{q^2\psi_h r_h^{2}E_h^{2}}{2\bigg(\frac{4n+3}{L^2}r_h^2+(4n+1)k-2^{2n-1}r_h^2 E_h^{2n+2}\bigg)^{2}}~. \label{Taylor6}
\end{eqnarray}
Moreover the above boundary conditions imply that the Dirichlet boundary condition $\psi(r=L)=0$ is satisfied by the test scalar field $\psi$ without imposing any constraint on other quantities $\eta(r), \xi(r)$ and $A(r)$ \cite{ref:Oscar2017}.

By numerically integrating the coupled differential equations (\ref{static1})-(\ref{static4}) as well as applying the above series expansions as initial conditions evaluated at
$r=r_h+\epsilon$ ($r_{h}=1,~\epsilon\sim10^{-15}-10^{-10}$) to circumvent singularity on the
event horizon, we are able to extract the static hairy BH solutions. Note that
henceforward we just focus on flat and spherical geometries ($k=0,~+1$)
since we have already found that for $k=-1$, the condition of small BH is violated.

\begin{figure}[!ht]
\includegraphics[width=8.5cm,height=5cm]{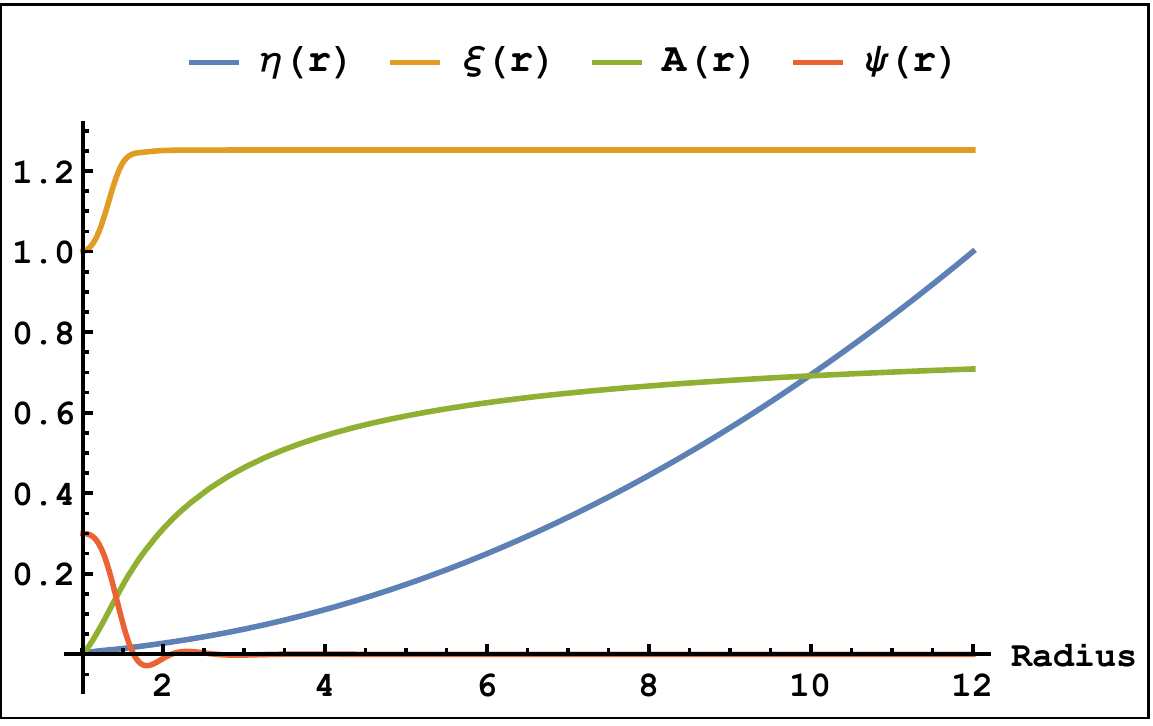}
\caption{A sample plot for two laps functions $\eta,~\xi$
and two matter fields $A,\psi$ for a particular static BH solution
with flat topology and conformal parameter $n=2$. The numerical values used are $q=0.7,~A'_{h}=0.2,~\psi_h=0.3$
with AdS radius $L=12$. There is no numerical solution for static BH with conformal parameter $n=1$.}
\label{flat}
\end{figure}

\begin{figure}[!ht]
\includegraphics[width=8cm,height=5cm]{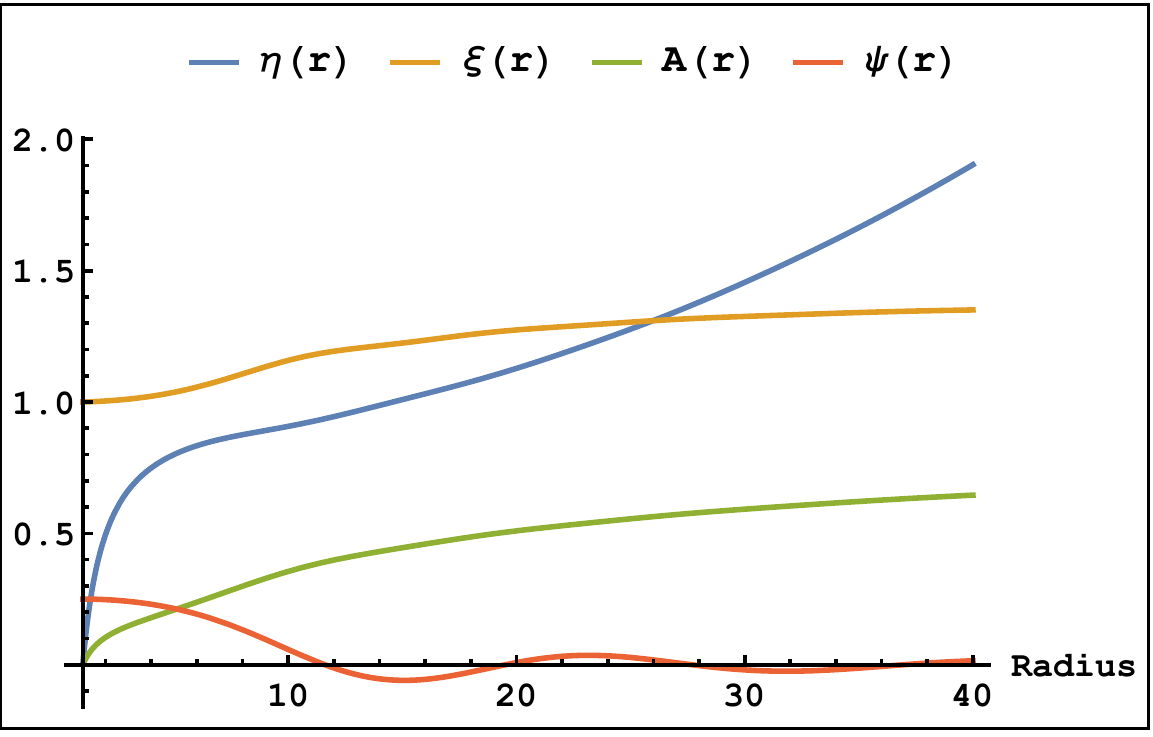}
\includegraphics[width=8cm,height=5cm]{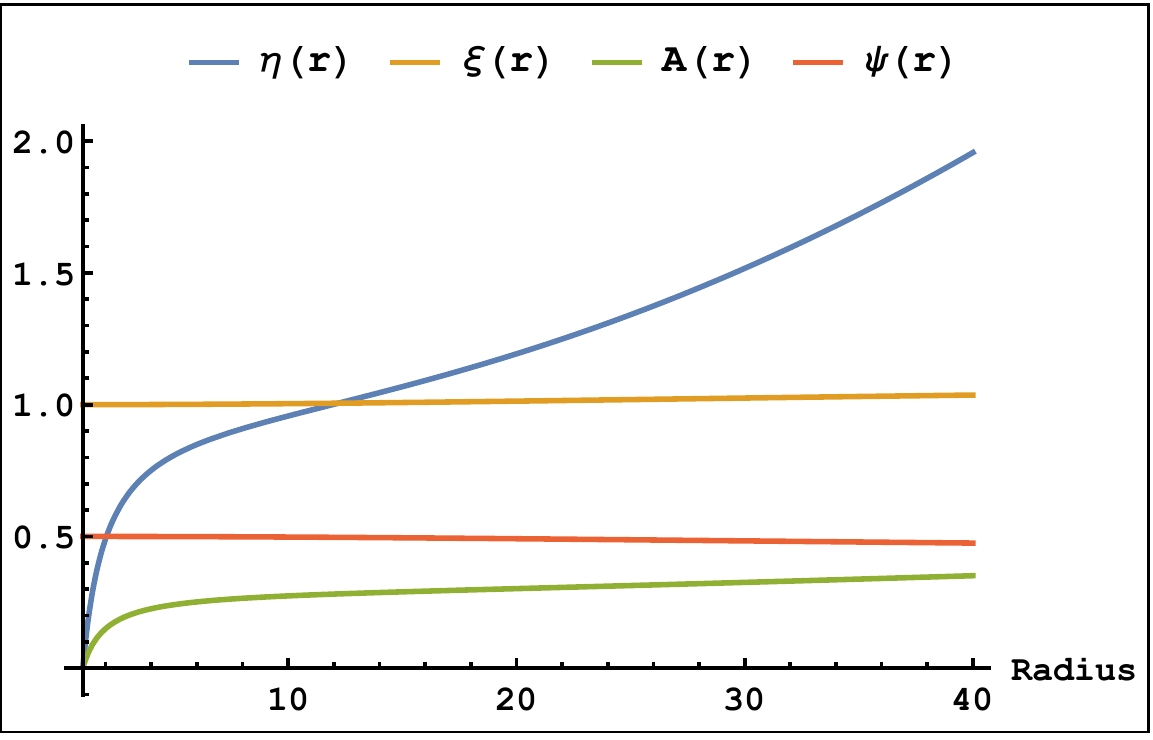}
\includegraphics[width=8cm,height=5cm]{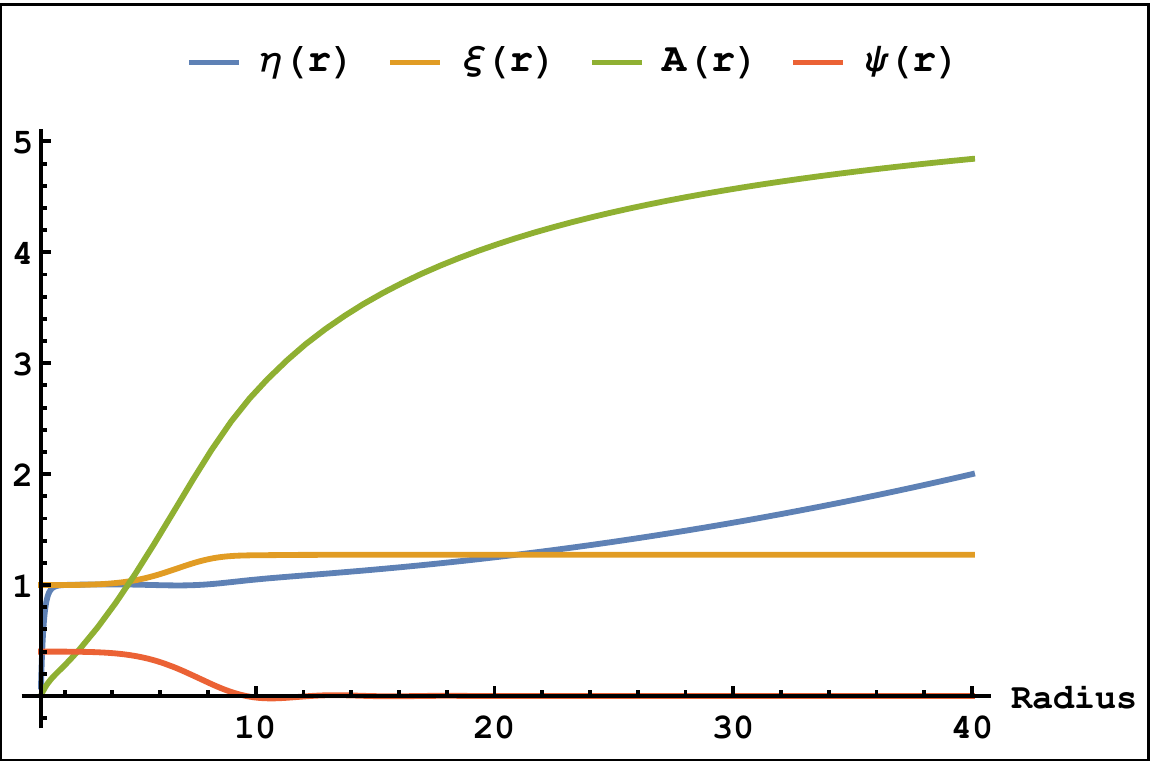}
\includegraphics[width=8cm,height=5cm]{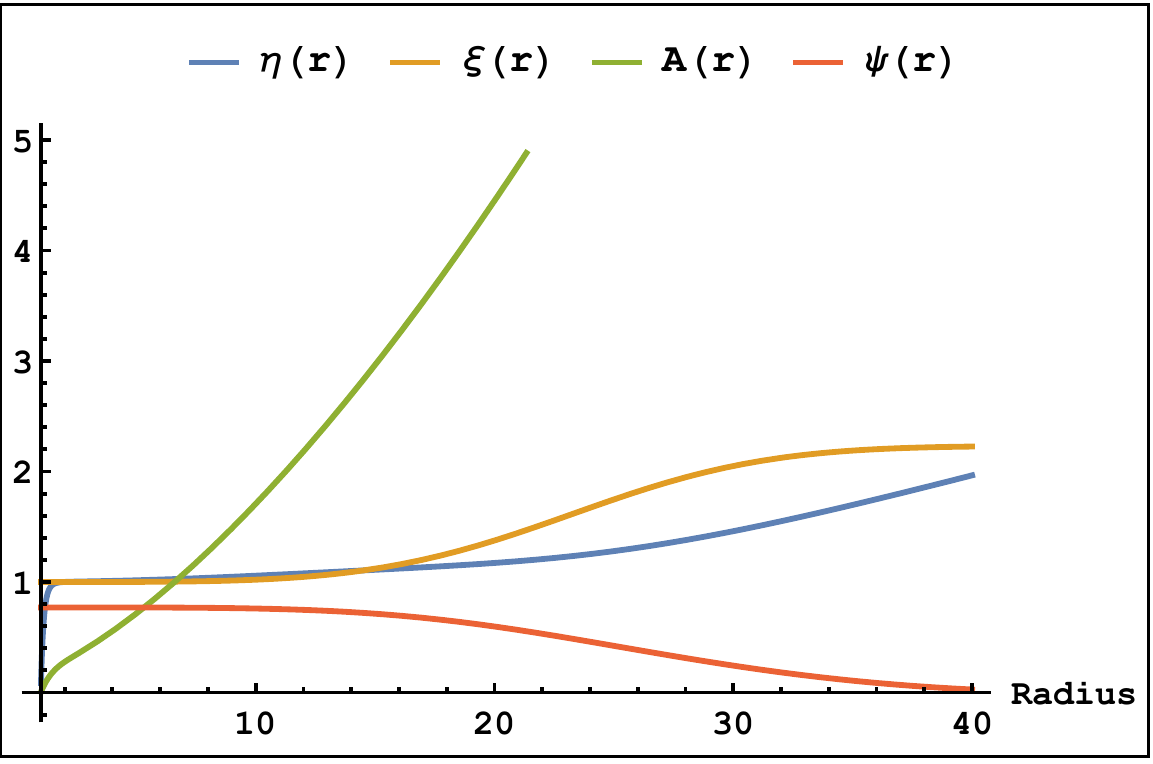}
\caption{A sample plots of the laps functions $\eta,~\xi$ and two
matter functions $A,\psi$ for a particular static BH solution with spherical geometrical geometry
and conformal parameters: $n=0$ (top row) and $n=2$ (down row). In the top row, we have used numerical values $q=0.9,~A'_{h}=0.2,~\psi_h=0.25$ and $q=0.06,~A'_{h}=0.3,~\psi_h=0.5$ for left and right
panels, respectively. In the bottom row, we have used numerical values
$q=0.6,~A'_{h}=0.5,~\psi_h=0.4$ and $q=0.055,~A'_{h}=0.55,~\psi_h=0.8$ for left and right
panels, respectively. The value of AdS radius is taken as $L=40$. There is no possibility of numerical
static BH solution for the conformal parameter $n=1$. }
\label{close}
\end{figure}

Let us first present the solutions
for variables\footnote{Concerning the far field behavior of variables $\eta(r),~\xi(r),~A(r)$, it should be noted that due to lack of an asymptotically flat BH solution, such solutions
include non trivial scalar hair \cite{ref:BekensteinI1972, ref:BekensteinII1972} so that it
is not necessity for these three variables to have finite values as $r\rightarrow\infty$.}
$\eta(r),~\xi(r),~A(r),~\psi(r)$ in Figs. \ref{flat} and \ref{close} for $k=0$
and $k=+1$, respectively. As can be seen, we are generally dealing with two
types solutions: single and multiple radial nodes for the scalar field $\psi(r)$. From
a stability perspective, a solution with a single radial node at the AdS boundary is desirable in
the sense that it addresses a scalar field solution enclosed between the horizon and AdS boundary
which, due to being in the ground state, enjoys the classical and quantum mechanical stability.
It should be stressed that although a single node scalar field solution can be a good candidate
to provide a stable hairy BH, further investigation in the form of spherically
symmetric perturbations is also needed. The solutions with more radial nodes represent the scalar
field in exited states (depending on the number of nodes) which is expected to be classically unstable.
Note that even in the event of classical stability of these exited solutions, there is yet the probability of their
transition to the ground state via quantum tunneling \cite{ref:1003}. As a result,
the scalar field solutions with more radial nodes are not prone to displaying
a stable hairy BH. The four solutions shown in Figs. \ref{flat}
and \ref{close} have been obtained for even conformal parameters $n=0$ or $n=2$ so that for
$n=1$,  there is no possibility of a numerical static BH solution.

Fig. \ref{close} tells us that in higher dimensions allowed by CIMS for a BH with a closed
horizon topology the possibility of having  single node solutions may exist.
However, the relevant single node scalar field solution is not important because
it appears only for a very small scalar charge ($q=0.055$). This is while we
showed earlier that for the small values of the scalar charge, instability
in higher dimensions does not occur. In what follows, by focusing only
on the variable $\psi(r)$, we will consider other possibilities.

\begin{figure}[!ht]
\includegraphics[width=8.5cm,height=5cm]{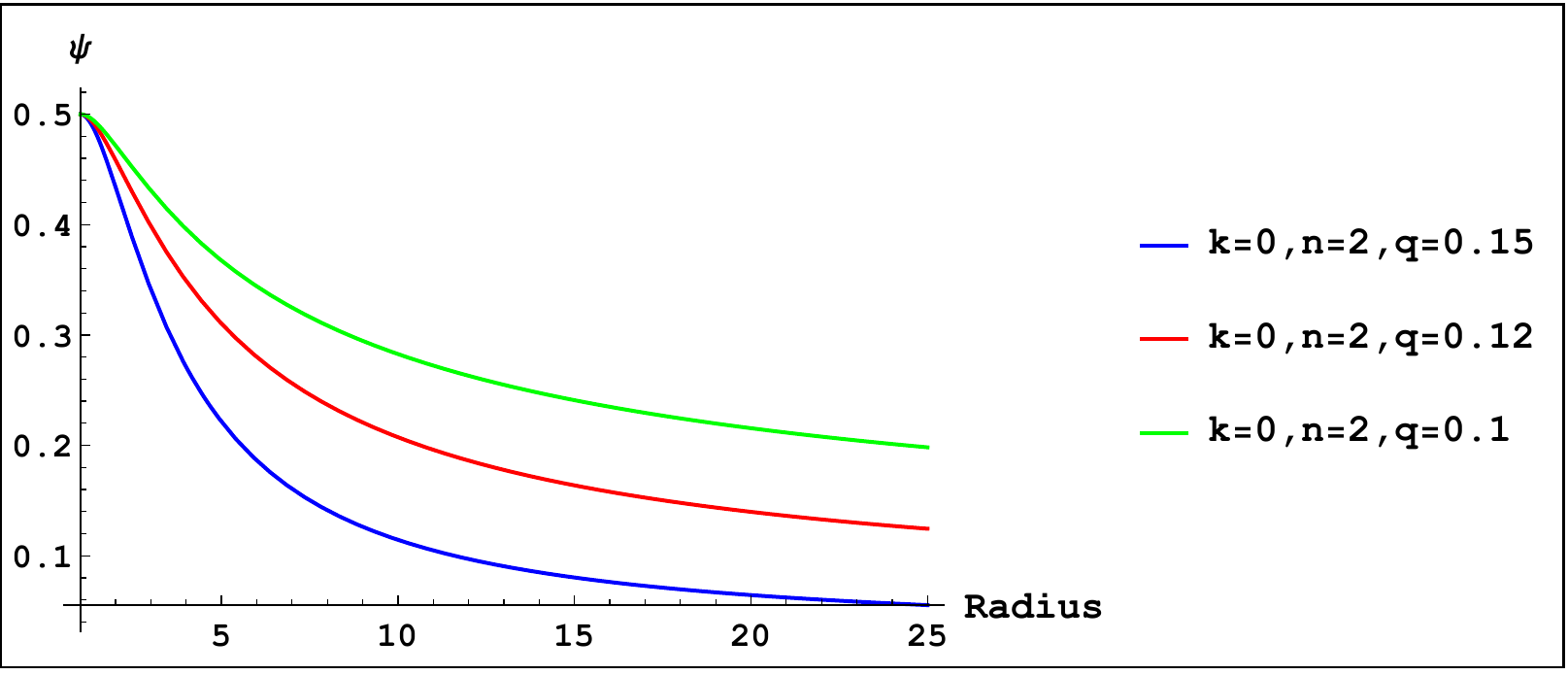}
\includegraphics[width=8.5cm,height=5cm]{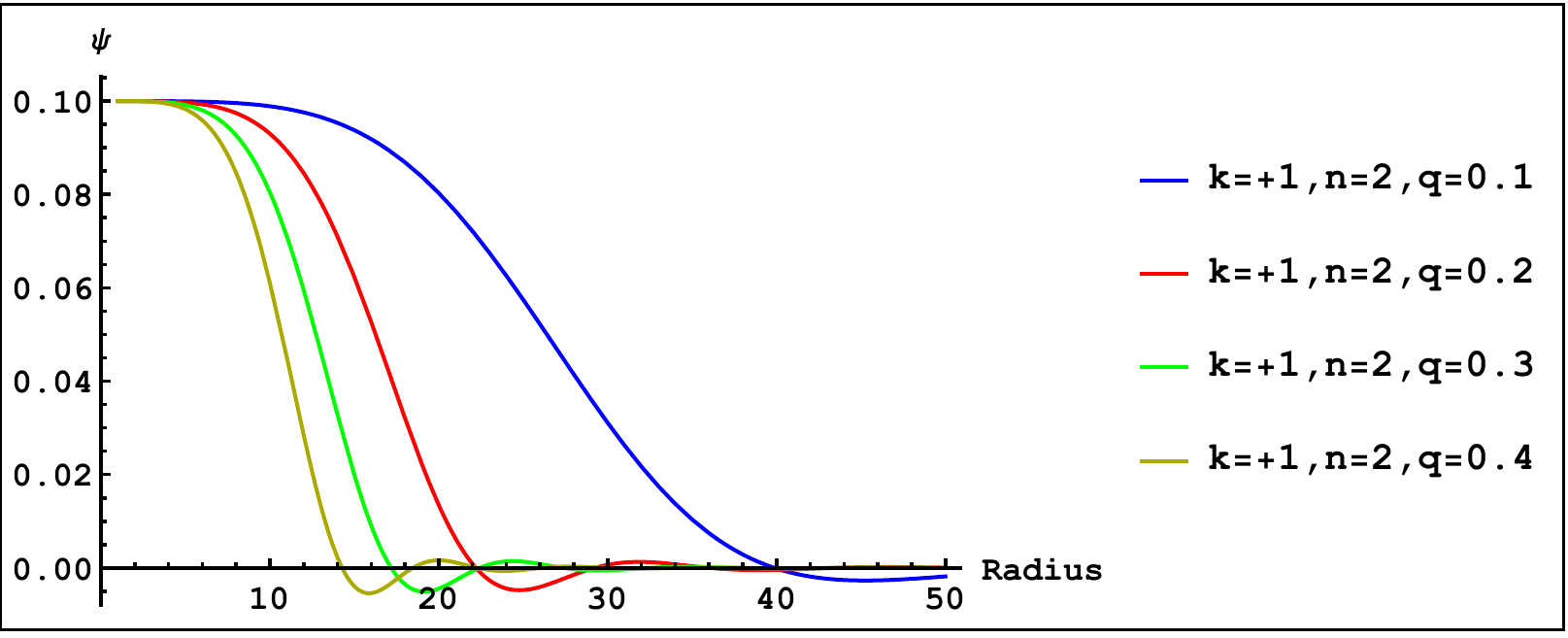}
\caption{A plot of the scalar field $\psi$ as a function
of radius for two horizon topologies: planar, left panel and
spherical, right panel. We used numerical values $L=10,~A'_{h}=0.5,~\psi_h=0.5$ and
$L=40,A'_{h}=0.3,~\psi_h=0.1$.}
\label{flatI}
\end{figure}

\begin{figure}[!ht]
\includegraphics[width=8.5cm,height=5cm]{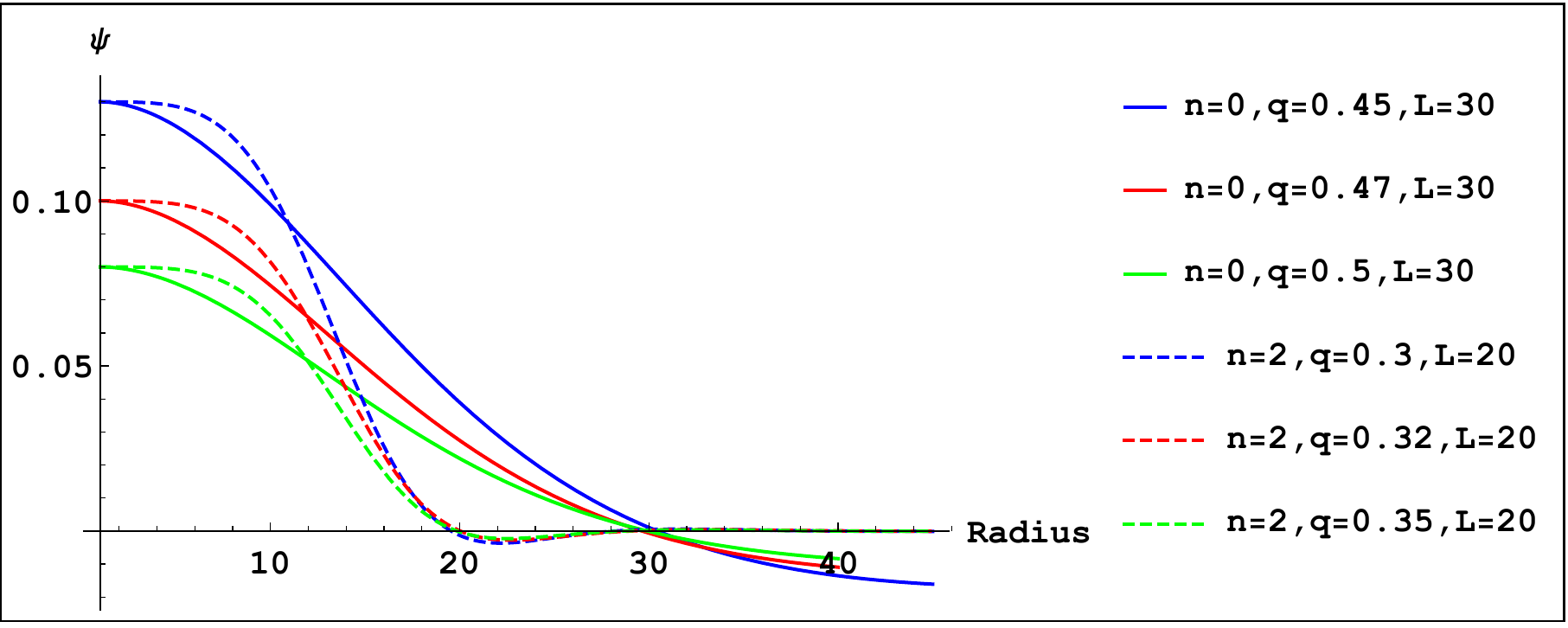}
\caption{A plot of the scalar field $\psi$ as a function
of radius for the case of spherical horizon topology ($k=+1$).
We used numerical values $A'_{h}=0.35,~\psi_h=0.13$.}
\label{closeI}
\end{figure}

Regarding the planar horizon topology, in Fig. \ref{flatI}, left panel, we find
an example of a scalar profile function $\psi(r)$ for which there is no
node in the AdS boundary. So, this type of profile cannot form a scalar hairy BH configuration
because it is not enclosed within the event horizon and AdS boundary. Generally, we find
that here there is no possibility of extracting a ground state scalar hairy BH configuration
for a horizon with flat geometry. However, for a spherical horizon topology we are dealing
with more profiles. As shown in Fig. \ref{flatI}, right panel, we see both the ground state
as well as exited scalar hairy BH solutions. It is clear that except for small scalar charges, e.g. $q=0.1$, we may see a ground state scalar hairy BH solution which, according to the above  reasoning, are of no interest.  In Fig. \ref{closeI} we also show that it is possible to have
different scalar hairy BH configurations which share the same AdS boundary. Here, at first glance, the idea may come into one's mind that the geometry of the event horizon for $k=+1$ and conformal parameter $n=2$ leads to ground state scalar hairy BH solutions. Recalling $L=\sqrt{-\frac{(d-1)(d-2)}{2\Lambda}}$, it is expected
that the AdS radius in allowed higher dimensions to be larger than that in 4-dimensions, that is $n=0$. This is while, the relevant figure obviously illustrates that the first node of the scalar hairy solution with conformal parameter $n=2$ is located within a radius smaller than that when $n=0$. As a result, the scalar hairy BH solutions shown here cannot truly address the ground state since before
the actual AdS radius is reached, there is at least one node present. Our numerical analyses do not address any scalar hairy BH solution for which its first node locates in an AdS radius larger than what is expected for the case when $n=0$. Overall, in spite of the existence of non-trivial scalar hairy BH solutions with planar and spherical symmetry horizons in solution space, it seems that such solutions are not able to form a scalar hairy BH
and the system still remains in its primary status \emph{i.e.,} RN-AdS background.
\section{Concluding remarks }\label{DC}
It is an established fact that the embedding of a charged BH in
an asymptotically AdS space-time may result in instability
due to the role played by the timelike boundary of space-time acting as
a natural reflecting boundary, the so-called Dirichlet wall.
The nature of this instability is superradiant if one
chooses the gauge $C=0$, since in this gauge the superradiance appears
for a particular set of states
with certain boundary conditions at infinity. However, in the gauge $C\neq0$
imposed directly by AdS-RN, we have encountered a superconducting-like instability
for a set of states that obey different boundary conditions at infinity. We note that
both sets of states are mapped into each other by means of
$U(1)$ gauge transformations which do not vanish at infinity, implying that they are different states.
This means that the instabilities are related, while addressing different physical phenomena.

In this paper, with the aim of investigating the no-scalar hair paradigm in higher
dimensions, we have followed the possibility of occurrence of instability
in a $d$-dimensional topological RN-AdS-small BH arising from Einstein-conformally
invariant Maxwell gravity.  In fact, the conformal invariance feature of the source does not allow arbitrary dimensions higher than four, but allows dimensions which are multiples of four, that is
$d=8,12,...$, equivalent to the conformal parameters $n=1,2,...$.

Employing a charged massless scalar perturbing test field with small amplitude,
we have applied a numerical procedure to investigate instability
at the level of linear approximation. Within the context of small black holes with $r_{h}\ll L$, the numerical analyses cannot tell us much about  instability for $k=-1$. In contrast,  for two modes $k=0,+1$ it was shown that by increasing the scalar
charge there is a possibility of higher dimensional
instability, see Fig. \ref{Super}. Next, by leaving linear approximation we have numerically extracted a family of static BH solutions, including a charged massless scalar-field
hair which is enclosed between the event horizon (with planar and spherical
horizon topologies) and the natural AdS boundary. Interestingly, for even
conformal parameters, in particular $n=2$, there are static solutions
with scalar profiles $\psi(r)$ including single and multiple radial
nodes which respectively address  the ground and excited states with classical stability/istability.
We found that for the relevant BH configuration with planar
horizon topology there is no possibility of extracting a ground state scalar hairy
solution, see the left panel in Fig. (\ref{flatI}), while for its spherical counterpart, there is. Generally, for BH configurations with spherical horizon topology with two configurations there is a chance to have ground state scalar hairy profiles; for small scalar charge (right panel in Fig. \ref{flatI}) and also for AdS length smaller than what we expect from the standard case $n=0$. However, none of these two cases are solid candidates to be identified as the ground state scalar hairy solutions. The former, as can be seen from the numerical solution shown in Fig. \ref{Super}, suggests that instability will not occur for small values of the scalar charge. The latter is due to what we expect to see as the first node at the AdS at distances larger than that for $n=0$, contrary to what is observed in Fig. \ref{closeI}. So, in spite of the existence of non-trivial scalar hairy BH solutions with  planar and spherical horizon symmetries, because of their weak probability of occurrence, the system has more chance to remain in RN-AdS background. As a consequence, no-scalar hair theorem is supported in the sense that the possibility of interaction of a scalar field with a RN-AdS BH with the consequence of the emergence of a new quantity such as mass and charge, does not exist.

\section{Acknowledgments}
We are grateful to Vitor Cardoso for reading the manuscript and useful comments. The work of M. Khodadi is supported by the
Research Institute for Astronomy and Astrophysics of
Maragha (RIAAM) under research project No. 1/6025-60. M. Honardoost would like to thank Iran National Science Foundation (INSF) and the Research Council of Shahid Beheshti University for financial support.
We would also like to thank the anonymous referee for insightful and constructive comments.

\end{document}